\documentclass[12pt]{article}
\usepackage{graphicx}
\usepackage{psfrag}

\def\tata{Tata Institute of Fundamental Research\\
Homi Bhabha Road, Mumbai 400 005, INDIA}
\def\support{\footnote{Work supported in parts by the Department of Atomic Energy
          and the Department of Science and Technology of India.}}

\def\Title#1{\begin{center} {\Large #1 } \end{center}}
\def\Author#1{\begin{center}{ \sc #1} \end{center}}
\def\Address#1{\begin{center}{ \it #1} \end{center}}

\newenvironment{Abstract}{\begin{quotation}  }{\end{quotation}}
\newenvironment{Presented}{\begin{quotation} \begin{center} 
             PRESENTED AT\end{center}\bigskip 
      \begin{center}\begin{large}}{\end{large}\end{center} \end{quotation}}
\def\Acknowledgements{\bigskip  \bigskip \begin{center} \begin{large}
             \bf ACKNOWLEDGEMENTS \end{large}\end{center}}

\def\beq{\begin{equation}}
\def\eeq#1{\label{#1}\end{equation}}
\def\eeqn{\end{equation}}

\def\beqa{\begin{eqnarray}}
\def\eeqa#1{\label{#1}\end{eqnarray}}
\def\eeqan{\end{eqnarray}}

\let\bar=\overbar

\def\Dslash{\not{\hbox{\kern-4pt $D$}}}
\def\dslash{\not{\hbox{\kern-2pt $\del$}}}

\def\msb{{\bar{\ssstyle M \kern -1pt S}}}

\RequirePackage{xspace}
\def\CP     {\ensuremath{C\!P}\xspace}
\def\qbar  {\ensuremath{\overline q}\xspace}
\def\cbar  {\ensuremath{\overline c}\xspace}
\def\Bbar   {\kern 0.18em\overline{\kern -0.18em B}{}\xspace}
\def\Dbar   {\kern 0.2em\overline{\kern -0.2em D}{}\xspace}

\mathchardef\Upsilon="7107
\def\Y#1S{\ensuremath{\Upsilon{(#1S)}}\xspace}

\def\mbc    {\mbox{$M_{\rm bc}$}\xspace}
\def\DeltaE {\mbox{$\Delta E$}\xspace}
\def\invfb  {\ensuremath{\mbox{\,fb}^{-1}}\xspace}
\newcommand {\mev}{\ensuremath{{\mathrm{\,Me\kern -0.1em V}}}\xspace}
\newcommand {\gev}{\ensuremath{{\mathrm{\,Ge\kern -0.1em V}}}\xspace}
\newcommand {\gevcc}{\ensuremath{{\mathrm{\,Ge\kern -0.1em V\!/}c^2}}\xspace}

\begin{document}
\begin{titlepage}

\vfill
\Title{Exotic, LFV and LNV Decays at the $B$ Factories}
\vfill
\Author{ Gagan B. Mohanty\support}
\Address{\tata}
\vfill
\begin{Abstract}
I review the latest results on exotic, lepton flavor violating (LFV) and
lepton number violating (LNV) decays of the $B$, $D$ mesons and the $\tau$
leptons, obtained at the two $B$-factory experiments, Belle and BaBar.
Where appropriate, results from other experiments are also described.
\end{Abstract}
\vfill
\begin{Presented}
6th International Workshop on the CKM Unitarity Triangle\\
University of Warwick, UK, September 6--10, 2010
\end{Presented}
\vfill
\end{titlepage}
\def\thefootnote{\fnsymbol{footnote}}
\setcounter{footnote}{0}

\section{Introduction}

The two $B$-factory experiments, Belle~\cite{belle} at KEK, Japan and
BaBar~\cite{babar} at SLAC, USA have played a key role in verifying
$\CP$ violation mechanism in the standard model (SM), which is
attributed to an irreducible phase that appears in the $3\times 3$
quark-flavor mixing matrix, known as the Cabibbo-Kobayashi-Maskawa
(CKM) matrix~\cite{ckm}. The focus of these experiments is now shifted
to search for new physics through a systematic investigation of decay
processes that are either heavily suppressed or forbidden within the SM.
For instance, decays that violate lepton flavor or lepton number -- two
good quantum numbers of the SM -- will provide a clean signature of
physics beyond the SM, if detected. The above strategy is complementary
to the direct search methods of the energy-frontier experiments at the
Tevatron and the LHC. In these proceedings, I summarize latest results
on exotic, lepton flavor violating (LFV) and lepton number violating
(LNV) decays of the $B$, $D$ mesons and the $\tau$ leptons, obtained
using the $e^+e^-$ collision data collected with Belle and BaBar.

\section{Search for the Decay {\boldmath $B^0\to\gamma\gamma$}}

\begin{figure}[htb]
\centering
\includegraphics[width=0.7\textwidth]{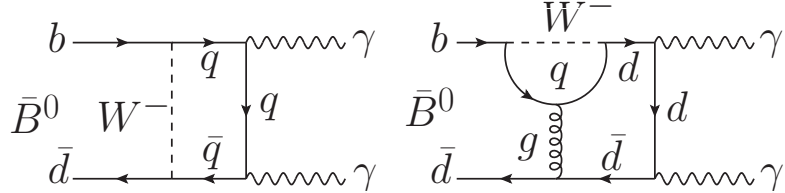}
\caption{Feynman diagrams for $B^0\to\gamma\gamma$. By replacing
the $d$ with a $s$ quark, one can get the contributing diagrams
for $B^0_s\to\gamma\gamma$. The symbol $q$ represents a $u$, $c$,
or $t$ quark.}
\label{fig:btogg1}
\end{figure}
In the SM, the decay $B^0\to\gamma\gamma$ proceeds through flavor
changing neutral current (FCNC) transition involving electroweak
loop diagrams, as shown in Fig.~\ref{fig:btogg1}. This decay is
suppressed with respect to $B^0_s\to\gamma\gamma$ by the CKM
factors $(|V_{td}/V_{ts}|^2\sim 0.04)$. The SM prediction for
the decay branching fraction is $\left(3.1^{+6.4}_{-1.6}\right)
\times10^{-8}$~\cite{bosch}. Potential new physics contributions,
{\it e.g.}, extended Higgs sector~\cite{aliev} or supersymmetry
with broken $R$-parity~\cite{gemintern} can significantly enhance
the decay rate.

BaBar has searched for the decay $B^0\to\gamma\gamma$~\cite{babargg}
using a data sample of $452\times 10^6$ $B\Bbar$ pairs collected at
the $\Y4S$ resonance. Signal events are selected using two kinematic
variables: the beam-energy-constrained $B$ mass $\mbc=\sqrt{s/4-
{\vec p}^{\,2}_{B}}$, and the difference $\DeltaE$ between the
center-of-mass (CM) energy of the $B$ candidate and $\sqrt{s}/2$,
where $\sqrt{s}$ is the total CM energy and ${\vec p}_B$ is the $B$
momentum in the CM frame. A two-dimensional unbinned maximum
likelihood fit to the $\mbc$--$\DeltaE$ distributions of $1679$
candidate events yields $21^{+13}_{-12}$ signal events. In absence
of a statistically compelling signal (significance is $1.9$ standard
deviations), a $90\%$ confidence level (CL) upper limit is calculated
for the branching fraction. The result, ${\cal B}(B^0\to\gamma\gamma)
<3.3\times 10^{-7}$, is nearly a factor of two below the best previous
upper limit from Belle~\cite{bellegg1}. Belle has also provided a
first upper limit~\cite{bellegg2} on the branching fraction of
$B^0_s\to\gamma\gamma$ $(8.7\times 10^{-6})$ using $23.6\invfb$ $\Y5S$
data. These limits allow further constraints on the new physics models.

\section{Search for {\boldmath $B^+\to D^-\ell^+\ell^{\prime+}$}}

\begin{figure}[htb]
\centering
\psfrag{L}{\boldmath $\ell^+$}
\psfrag{G}{\boldmath $\ell^{\prime +}$}
\includegraphics[width=0.6\textwidth]{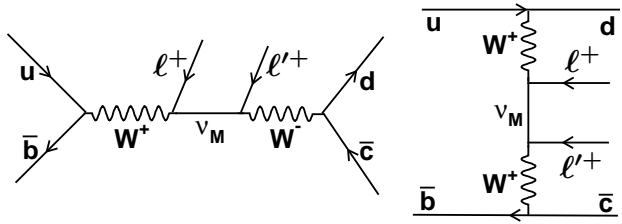}
\caption{Diagrams involving massive Majorana neutrinos
($\nu_{\rm M}$) that contribute to $B^+\to D^-\ell^+\ell^{\prime +}$
(plus the same diagrams with leptons exchanged if $\ell=\ell^\prime$).}
\label{fig:btodll}
\end{figure}

With the evidence for neutrino oscillation~\cite{garcia} being
on a firm footing, which points to a nonzero neutrino mass, one
of the fundamental questions in particle physics today is whether
neutrinos are standard Dirac or Majorana (antiparticle to
itself) fermions. If they are of the latter kind, the lepton
number would be no more a good quantum number in addition to
the lepton flavor, which seems to have already violated in
neutrino oscillation. This would lead to both LFV and LNV
decays $B^+\to h^-\ell^+\ell^{\prime+}$, where $h$ is a $D$,
$K$, or $\pi$ and $\ell\ell^\prime$ are $ee$, $e\mu$, or
$\mu\mu$. As an example, in Fig.~\ref{fig:btodll} we show the
contributing diagrams for $B^+\to D^-\ell^+\ell^{\prime +}$.
Given the tiny neutrino mass, the expected event rates for
these processes are many orders of magnitude below current
experimental sensitivity. Therefore, any observation would
be clear evidence for new physics, {\it e.g.}, model involving
a heavy Majorana neutrino of mass in the range $2$--$4\gevcc$~\cite{atre}.

Belle has conducted a first search of the decay $B^+\to D^-
\ell^+\ell^{\prime+}$~\cite{btodll}, where the $D^-$ decays to $K^+\pi^-\pi^-$
using $772\times 10^6$ $B\Bbar$ decays. First, a likelihood
method, mostly based on event shape variables, is designed
to suppress the $e^+e^-\to q\qbar$ $(q=u,d,s,c)$ continuum
background. Then, a counting analysis is performed where the
number of background events expected in the signal region is
evaluated from a $\mbc$--$\DeltaE$ sideband. The results are
presented in Table~\ref{tab:btodll}, where $90\%$ CL upper
limits on the branching fractions are also quoted since the
data are consistent with the background expectations. BaBar
has significantly improved~\cite{babar-hll} the previous
limits from CLEO~\cite{cleo-hll} for the LFV decays $B\to
K e^{\pm}\mu^{\mp}$, $B\to K^\star e^{\pm}\mu^{\mp}$, and
$B\to\pi e^{\pm}\mu^{\mp}$. The results are $5.1\times
10^{-7}$, $3.8\times 10^{-8}$, and $9.2\times 10^{-8}$,
respectively. For other $B\to h\ell^{\pm}\ell^{\prime \mp}$ and
$B^+\to h^-\ell^{+}\ell^{\prime +}$ decays, CLEO~\cite{cleo-hll}
has the world's best limit.
\begin{table}[!htb]
\begin{center}
\begin{tabular}{c|c|c|c|c}
\hline\hline
Decay mode & $\epsilon$ & $N_{\rm bkg}$ & $N_{\rm obs}$ & UL on ${\cal B}$\\
\hline
$B^+\to D^-e^+e^+$&$1.2\%$&$0.2\pm0.1$&$0$&$2.7\times 10^{-6}$\\
$B^+\to D^-e^+\mu^+$&$1.3\%$&$0.8\pm0.3$&$0$&$1.9\times 10^{-6}$\\
$B^+\to D^-\mu^+\mu^+$&$1.8\%$&$1.4\pm0.4$&$0$&$1.1\times 10^{-6}$\\
\hline\hline
\end{tabular}
\caption{Efficiency ($\epsilon$), expected background yield ($N_{\rm bkg}$),
number of data events ($N_{\rm obs}$), and $90\%$ CL upper limit on the
branching fractions for $B^+\to D^-\ell^+\ell^{\prime +}$.}
\label{tab:btodll}
\end{center}
\end{table}

\section{Search for Leptonic {\boldmath $D$} Decays}

\begin{figure}[htb]
\centering
\includegraphics[width=0.35\textwidth]{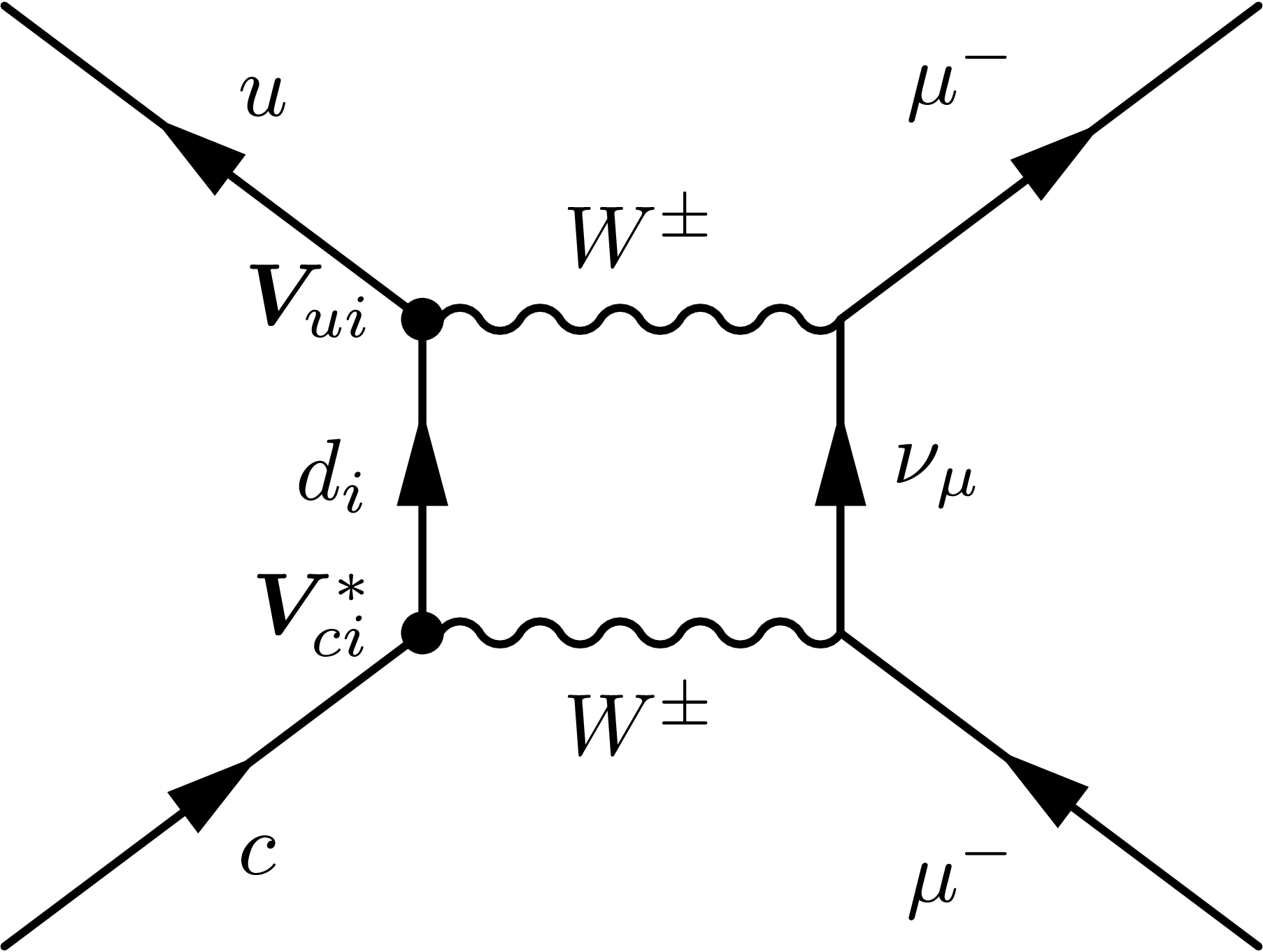}
\includegraphics[width=0.35\textwidth]{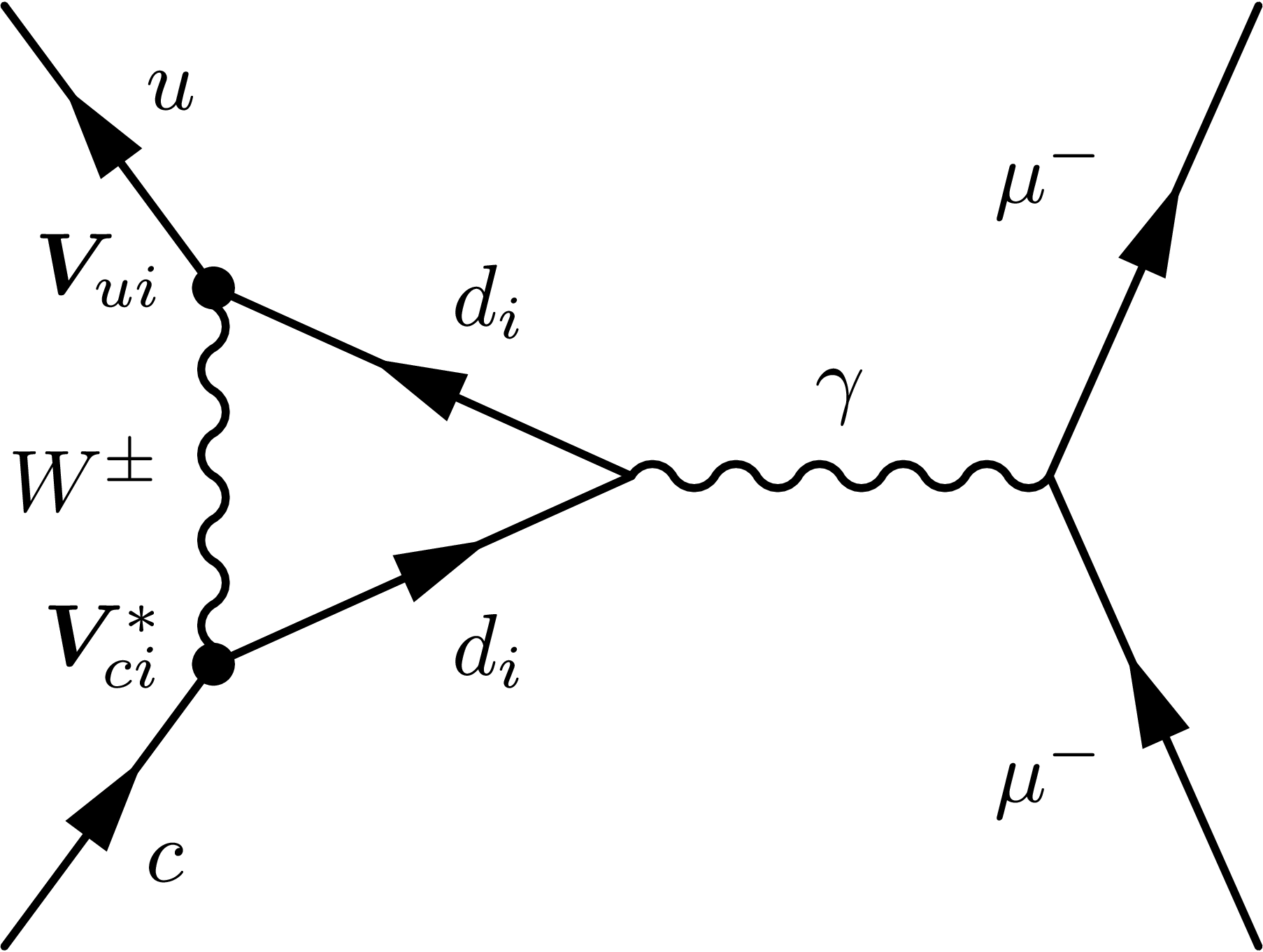}
\caption{Feynman diagrams for the $D^0\to\mu^+\mu^-$ decay.}
\label{fig:dtoll}
\end{figure}

The FCNC decays $D^0\to\ell^+\ell^-$ $(\ell=e/\mu)$ are highly
suppressed in the SM because of the GIM mechanism~\cite{gim}.
Figure~\ref{fig:dtoll} shows typical Feynman diagrams for $D^0\to
\mu^+\mu^-$. Potential new physics scenarios, such as $R$-parity
violating SUSY~\cite{golowich}, can raise their branching fraction
close to current experimental sensitivity. The LFV decay $D^0\to
\ell^+\ell^{\prime -}$ $(\ell\neq\ell^\prime)$ is SM forbidden,
but is possible in extensions of the SM, such as nondegenerate
neutrinos~\cite{burdman}. Any signal here, therefore, would be
a signal of new physics.

Belle has performed a search for $D^0\to\ell^+\ell^{\prime -}$
\cite{belle-dll} using $660\invfb$ data taken at, and $60\mev$
below, the $\Y4S$ peak. To suppress higher combinatorial
background associated with $D^0$ mesons coming from $B$ decays,
only those resulting from the decay $D^{\star +}\to D^0\pi^+$
in the process $e^+e^-\to c\cbar$ are considered. The measurement
is carried out with respect to a well measured and topologically
similar channel $D^0\to\pi^+\pi^-$, which allows cancellation
of the common systematic uncertainties. Candidate $D^0$ mesons
are identified using two kinematic variables: the invariant mass
of the $D^0$ daughters $M$, and the energy released in the
$D^{\star +}$ decay $q$. An optimal requirement on the size
of the signal region in $M$ and $q$ is imposed to minimize the
expected background in that region. Table~\ref{tab:dll}
summarizes the search result. There is no evidence for a
signal, and $90\%$ CL upper limits are set on the branching
fractions. These results improve with respect to the best
previous limits from BaBar~\cite{babar-dll} by a factor
of $9$ for $D^0\to\mu^+\mu^-$, $15$ for $D^0\to e^+e^-$,
and $3$ for $D^0\to e^{\pm}\mu^{\mp}$. A competitive limit
($2.1\times 10^{-7}$) for the channel $D^0\to\mu^+\mu^-$ also
exists from the CDF Collaboration~\cite{cdf-dll}. All these
results constrain the size of certain $R$-parity violating
couplings~\cite{golowich}, and strongly disfavor a leptoquark
contribution~\cite{dorsner} as the possible explanation for
the discrepancy between the expected~\cite{lattice} and observed~\cite{HFAG} $f(D^+_s)$ values.

\begin{table}[!htb]
\begin{center}
\begin{tabular}{c|c|c|c|c}
\hline\hline
Decay mode & $\epsilon[\%]$ & $N_{\rm bkg}$ & $N_{\rm obs}$ & UL on ${\cal B}$\\
\hline
$D^0\to\mu^+\mu^-$       &$7.02\pm 0.34$& $3.1\pm 0.1$&$2$&$1.4\times 10^{-7}$\\
$D^0\to e^+ e^-$         &$5.27\pm 0.32$& $1.7\pm 0.2$&$0$&$7.9\times 10^{-8}$\\
$D^0\to e^{\pm}\mu^{\mp}$&$6.24\pm 0.27$& $2.6\pm 0.2$&$3$&$2.6\times 10^{-7}$\\
\hline\hline
\end{tabular}
\caption{Efficiency ($\epsilon$), expected background yield
$(N_{\rm bkg})$, number of data events ($N_{\rm obs}$), and
$90\%$ CL upper limit on the branching fractions for
$D^0\to\ell^+\ell^{\prime -}$.}
\label{tab:dll}
\end{center}
\end{table}

\section{Lepton Flavor Violation in {\boldmath $\tau$} Decays}

LFV decays of charged leptons are expected to have negligible
rates in the SM even after including neutrino oscillation effects,
{\it e.g.}, ${\cal B}(\tau^-\to\mu^-\gamma)<10^{-54}$~\cite{pham}.
Therefore, it is impossible to observe these decays in current
experiments. However, many new physics scenarios, such as
supersymmetry~\cite{brignole} and large extra dimensions~\cite{chang}, predict enhanced
LFV decays with branching fractions being close to current
experimental sensitivity. In most of these models, $\tau$
leptons are expected to be strongly coupled and to have many
possible LFV decays due to their large mass. Therefore, LFV
$\tau$ decays provide an ideal probe for physics beyond the SM.

\begin{figure}[htb]
\centering
\includegraphics[width=0.75\textwidth]{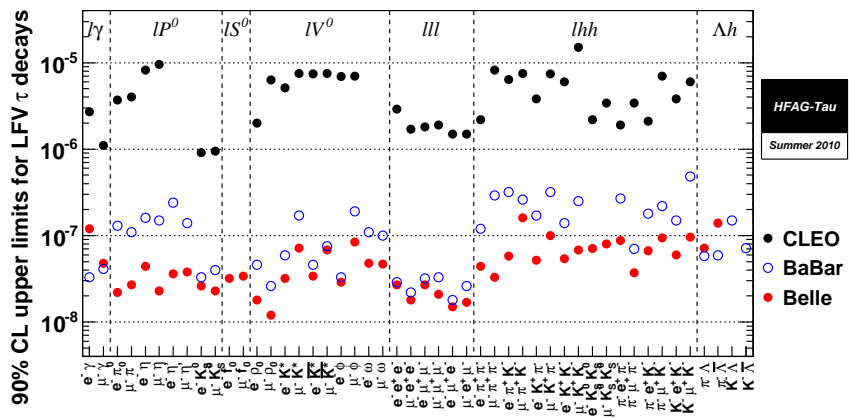}
\caption{Summary of searches for LFV $\tau$ decays~\cite{HFAG}.}
\label{fig:lfv-tau}
\end{figure}

Belle and BaBar have performed many intensive searches for
LFV $\tau$ decays, using $e^+e^-\to\tau^+\tau^-$ data samples
collected near the $\Y4S$ resonance. The analysis strategy is
similar for both the experiments. Starting with $\tau^+\tau^-$
events, the $\tau$ decaying into an LFV mode is called the
`signal side', while the other one (`tag side') is detected
via its decay into one charged particle with any additional
number of neutrals including neutrinos. Candidate events
surviving basic signal selection criteria are examined in
the two-dimensional space of the reconstructed mass of the
signal side ($M_{\rm sig}$), and the difference of the signal-side
$\tau$ energy from the beam energy in the CM frame ($\DeltaE_\tau$).
The signal region in the $M_{\rm sig}$--$\DeltaE_\tau$
plane remains blinded until the number of expected background
events in that region is evaluated. Finally, the latter is
compared with the number of data events to see whether there
is an excess in data (`observation') or the data are consistent
with the background hypothesis. No evidence for LFV $\tau$ decay
is found, and $90\%$ CL upper limits are set on the branching
fractions. Most of the upper limits are in the range
$10^{-7}$--$10^{-8}$ (see Fig.~\ref{fig:lfv-tau}), with the
most sensitive result coming from $\tau^-\to\mu^-\rho^0$.
These stringent limits can be used to constrain the parameter
space of various new physics models.

\section{Closing Remarks}

After accomplishing the major goal of their inceptions, which was to
establish the CKM framework as the source of $\CP$ violation in the
SM, Belle and BaBar have turned their attention to rare decays. Using
a large, clean data sample in conjunction with sophisticated analysis
methods, they are exploring decays that may not have even been
thought of at their beginning about a decade ago. The baton is being
gradually handed down to the next generation flavor experiment, LHCb
\cite{lhcb} at the LHC. With the proposed super flavor factories
\cite{superb} being well in sight, the future seems to be brighter
for experimenters at the luminosity frontier.

\Acknowledgements
I am grateful to B. Golob, M. Gorbahn, S. Robertson, Y. Sakai, and
K. Trabelsi for their important inputs during the preparation of my
talk and these proceedings.

\end{document}